\documentclass{raa}
\usepackage{graphicx,times}
\usepackage{natbib}
\usepackage{amssymb,amsmath}
\usepackage{subfigure}
\bibpunct{(}{)}{;}{a}{}{,}

\usepackage[a4paper=true,dvipdfm=true,pagebackref=true]{hyperref}
\hypersetup{pdftitle = The title of my PDF, pdfauthor = My name, pdfsubject= The subject, pdfkeywords = keyword1 keyword2 keyword3}
\hypersetup{colorlinks = true, linkcolor = green, anchorcolor = red, citecolor = blue, filecolor = red, pagecolor = red, urlcolor = red}
\begin{document}

\title{Supernova Neutrino in a Strangeon Star Model}

 \volnopage{Vol.0 (201x) No.0, 000--000}      
   \setcounter{page}{1}          
   
 \author{Mao Yuan\inst{1}, Jiguang Lu\inst{2}, Zhiliang Yang\inst{1}, Xiaoyu Lai\inst{3}, Renxin Xu\inst{2,4}}
 \institute{ Department of Astronomy, Beijing Normal University, Beijing 100875, China \\
  \and
  School of Physics, Peking University, Beijing 100871, China\\
  \and
  School of physics and Engineering, Hubei University of Education, Wuhan 430205, China\\
  \and
  Kavli Institute for Astronomy and Astrophysics at Peking University, Beijing 100871}

  \abstract{The neutrino burst detected during supernova SN1987A is explained in a strangeon star model, in which it is proposed that a pulsar-like compact object is composed of strangeons ({\em strangeon}: an abbreviation of ``strange nucleon'').
  A nascent strangeon star's initial internal energy is calculated, with the inclusion of pion excitation (energy around $10^{53}$\,erg, comparable to the gravitational binding energy of a collapsed core).
  A liquid-solid phase transition at temperature $\sim 1-2$ MeV may occur only a few ten-seconds after core-collapse, and the thermal evolution of strangeon star is then modeled.
  It is found that the neutrino burst observed from SN 1987A could be re-produced in such a cooling model.
  }
  \maketitle

  \section{Introduction}           
\label{sec1}

The state of dense baryonic matter compressed during supernova is not yet well understood because of the non-perturbative nature of the fundamental strong interaction, but it is popularly speculated that those compact stars are composed of nucleons (this kind of matter should actually be neutron rich because of the weak interaction, thus we usually call them as neutron stars). However, it has already been proposed that these compact stars could be composed of strangeons, formerly known as quark-clusters or strange clusters (\citealt{Xu+2003}). Strangeon is actually an abbreviation of ``strange nucleon'', in which the constituent quarks are of three flavors (up, down, and strange) rather than of two for nucleons.
Both normal nucleus and strangeon matter are self-bound by residual color interaction, so we may simply call a strangeon star a gigantic nucleus with strangeness.

Because of the massive (thus non-relativistic) nature of strangeons and the short-distance repulsion force between them (an analogy of the nuclear hard core), the equation of state of strangeon matter is very stiff (\citealt{Lai+etal+2009}) so that the observations of two-solar mass pulsars (\citealt{Demorest+etal+2010, Antoniadis+etal+2013}) could be naturally explained.
Strangeon matter would be solidified when its temperature is much lower than the residual interaction energy in-between (\citealt{Dai+etal+2011}), and pulsar glitches, with or without X-ray enhancement, could be understood in the
regime of starquake since the energy release depends
on spin frequency in solid strangeon star model (\citealt{Zhou+etal+2014}).
In addition, the quake-induced release of both gravitational and elastic energy could be meaningful for anomalous X-ray pulsars and soft gamma-ray repeaters (\citealt{Xu+Tao+etal+2006, Tong+etal+2016}). Because of the strangeness barrier on stellar surface, the optical/UV excess of X-ray dim isolated neutron star could then be understood by including the free-free emission from a strangeon star atmosphere (\citealt{wang+etal+2017}).  A strangeon star could be spontaneously magnetized due to ferromagnetic transition of electrons (\citealt{Lai+etal+2016a}), and some of small glitches could be the results of collisions between strangeon stars and strangeon nuggets (\citealt{Lai+etal+2016b}).
Despite these successes listed above, a general question arises: Is it possible to understand the neutrino burst observed during supernova 1987A in the regime of strangeon star?

This is the question we are attempting to answer in this paper.
Normal 2-flavor baryonic matter could be transformed into strangeon matter through strangeonization during a compression process.
Similar to the neutronization process of $e+p\rightarrow n+\nu_{e}$, a strangeonization process of $(u,d) \rightarrow (u,d,s)$ will also significantly kill off electrons and hence produce strange ``nucleons'', i.e., strangeons.
A strangeon is a cluster of quarks with quark number, $N_{\rm q}$ (probably 6, 9, 12 or 18).
Being different from a strange quark star (SQS), as mentioned above, a strangeon star (SS) could be converted to a solid star from a liquid one, with melting temperature $T_{\rm m} \sim$ MeV (\citealt{Dai+etal+2011}). Namely, after a phase transition the whole SS could be in a solid state during its cooling process.

A photon-driven mechanism would work for both SQS and SS (e.g., \citealt{Chen+etal+2007}), alleviating the difficulty of traditional neutrino-driven supernova (\citealt{Thompson+etal+2003}). Due to extremely high temperatures, significate neutrinos are radiated during a photon-driven supernova.
The total photon energy released could be as much as $\sim 10^{52}$ erg according to our calculations below, while neutrinos still take away almost all of the gravitational energy $\sim 10^{53}$ erg.
In contrast to the conventional neutrino-driven model, neutrinos are usually trapped in a nascent SS due to high opacity caused by coherent scattering off strangeons, that means that an SS's ``neutrinosphere'' could be of the same scale of that of the proto-star.
In this scenario, the neutrino emissivity of SS depends on the temperature of the whole nascent SS, rather than on the thin layer of a proto-star.
Is it possible to test the scenario through neutrino observation?
Luckily, in 1987 a neutrino burst in a core-collapse supernova SN 1987A was detected by 3 detectors, Kamiokande-\uppercase\expandafter{\romannumeral2} (\citealt{Hirata+etal+1987}), Irvine-Michigan-Brookhaven (IMB) (\citealt{Bionta+etal+1987}) and Baksan (\citealt{Alekseev+etal+1987}) almost at the same time. So far this is the only time that astronomers have observed neutrinos from new born compact objects. In this paper, we discover whether the cooling behaviour of SS can match the observation of SN1987A neutrino burst.

This paper consists of the following parts. The study of the whole thermal evolution of a new born SS is presented in \S~\ref{sec2}, which includes the calculations of the internal energy of new born SS with different masses in \S~\ref{sec2.1}, the radiation of the proto-SS in \S~\ref{sec2.2}, and the specific thermal evolution and phase transition in \S~\ref{sec2.3}. After the theoretical calculations we introduce the neutrino burst from SN 1987A and reproduce it with our model in \S~\ref{sec3}. Finally in \S~\ref{sec4} is the conclusions we have reached as well as some discussions.

  \section{Thermal evolution of a newborn Strangeon Star}
  \label{sec2}
Huge internal energy is stored in a newborn SS after collapse, and then the energy is released by photons and neutrinos. This process is dominated by neutrino radiation. During this cooling process a sharp drop in temperature leads to a phase transition of an SS. In this section, we make a rough calculation about this evolution process.
  \subsection{Internal energy}
  \label{sec2.1}

Different from hadron stars and hybrid stars bounded by gravity, an SS is a self-bounded object that bounded by residual color-interactions between strangeons. Correspondingly, equation of state, which is distinctly reflected in $M-R$ relations, varies in different models. The $M-R$ relations of gravitationally bound neutron stars have been proposed by many authors (\citealt{Prakash+1987, Prakash+etal+1988,  Akmal+Pandharipande+1997, Glendenning+Schaffner-Bielich+1999}), and the results showed that a more massive neutron star might correspond to a smaller radius. Generally, the mass higher than 2$M_{\odot} $ is difficult to explain in these models, but this is natural in SS model. The main reason is the different mass density gradient from stellar center to its surface, and the strangeon matter could have a stiff equation of state due to the strong coupling (\citealt{Guo+etal+2014}).

\begin{figure}
   \centering
   \includegraphics[width=10.0cm, angle=0]{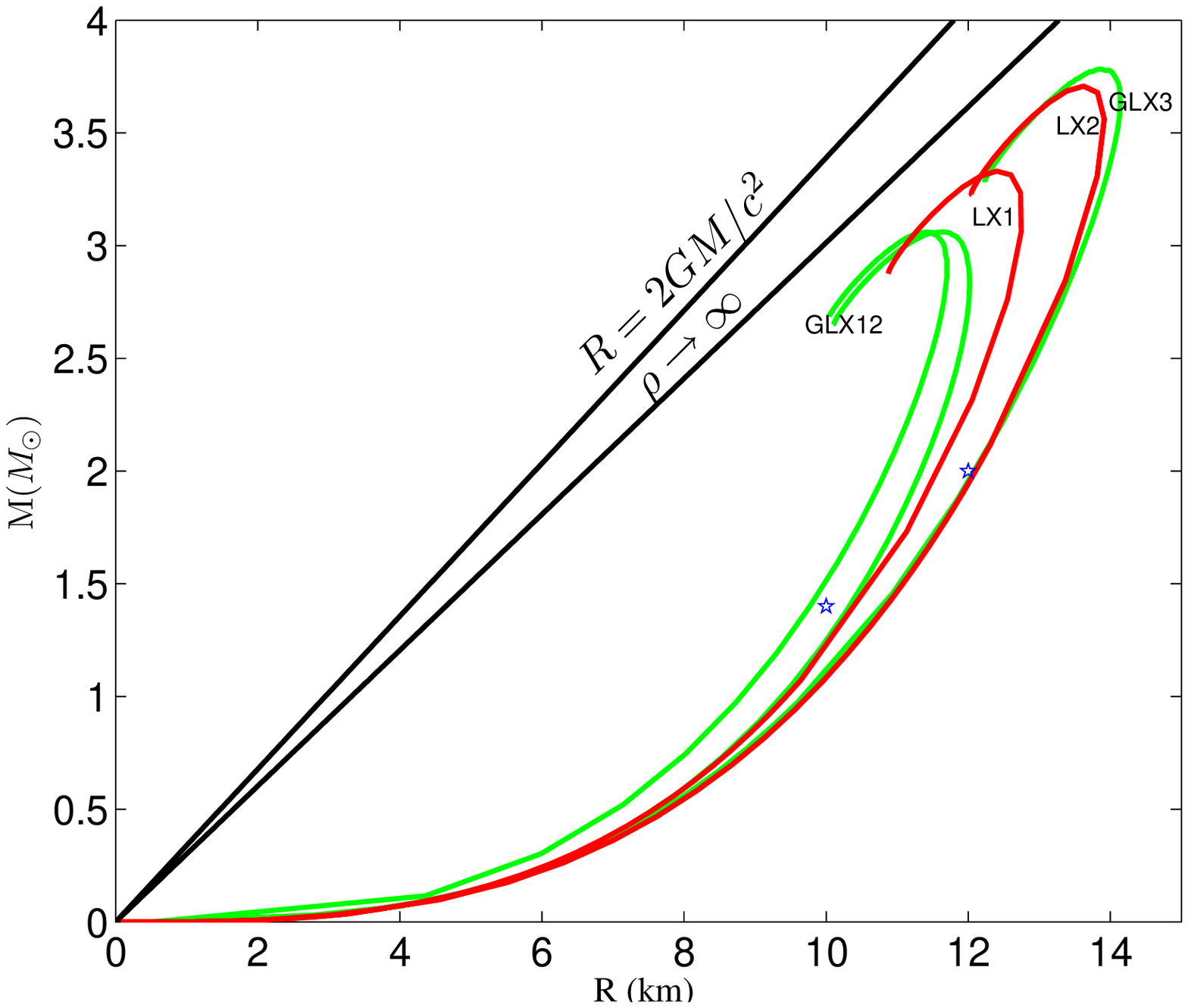}
   \caption{$M-R$ relations for SSs. The upper black lines show the general relativity and central density limit. GLX123 (\citealt{Guo+etal+2014}) and LX12 (\citealt{Lai+etal+2009, Lai+etal+2013}) represent the theoretical $M-R$ relations for SSs. It is clear that the SS model can support a pulsar-like star with a mass of more than 2$M_{\odot}$. Since the compact remnant of SN 1987A is still unobservable (\citealt{Manchester+Peterson+1996, Manchester+2007}), no further information could be obtained about its size and mass. In this paper, we parameterize the mass of a new-born strangeon star to be $1.4M_\odot, 2M_\odot$ which are shown as the pentalpha, with corresponding radii of $10$km, $12$km.}
   \label{Fig1}
   \end{figure}

Mass density $\rho$ consists of rest-mass density and energy density, for an SS, which reads,
\begin{equation}
\rho=n_{\rm s}(N_{\rm q}m_{0}+E/c^{2}) \label{eq1},
\end{equation}
where $n_{\rm s}$ is the number density of strangeons, $m_{0}$ is the constituent quark mass, and $N_{\rm q}$ is a free parameter to be about the number of quarks in each strangeon. As mentioned in \S~\ref{sec1}, we take $N_{q}=6\sim18$ for each strangeon. Energy density, $E$, in Eq.~\ref{eq1} contributes little to the mass density (\citealt{Guo+etal+2014}) in our model, so it is ignored in the following calculations but significant for the equation of state. On the surface, the density could approximate of the rest-mass density $\rho_{s}=n_{s}N_{q}m_{0}$. From Tolman-Oppenheimer-Volko (TOV) equations, equation of state for SSs can be derived (\citealt{Lai+etal+2009, Lai+etal+2013, Guo+etal+2014, Li+etal+2015}), and $M-R$ relations can be obtained, as Fig. \ref{Fig1} shows. We take two sets of different parameters (the pentalphas showed in Fig.\ref{Fig1}) of SSs in following calculations for indication, including the typical 1.4$M_{\odot}$ and the other of massive pulsar (2$M_{\odot}$).

When a proto-compact star is formed in the iron core of an evolved massive star, it goes through a transition process from gravitational energy (or the binding energy $E_{\rm bind}$) to a star's internal energy which is around $10^{53}$ erg. The gravitational energy would be stored in SS matter as a form of initial thermal energy (or internal energy) during strangeonization process. Consequentially, the initial temperature of a proto-SS is extremely high with several $10^{11}$ K, just like that of a proto-neutron star. In addition to strangeons, degrees of freedom in a proto-SS are uncertain with such high temperature and high density. Migdal did a lot of research on the phase transition of baryons in super-dense stars (\citealt{Migdal+1972, Migdal+N+1973, Migdal+JETP+1973, Migdal+etal+1973}). It indicated that new degrees of freedom, mesons, could be excited due to vacuum instability in a super-dense object. Based on Migdal's arguments, we suggest that a huge number of pions (including $\pi^{0}, \pi^{+}, \pi^{-}$) would be excited in a newborn SS.

Phenomenologically, pions (with mass $m_{\pi^{0}}=134.98$ MeV, $m_{\pi^{\pm}}=139.57$ MeV ) are the lightest carriers of residual strong interactions between strangeons, so they can be excited more easily than other mesons. Other freedom degrees could be leptons (e.g., neutrinos and positrons) and photons, as well as kinematical oscillation of strangeons. All of these components share the gravitational binding energy and store it as internal energy, which is
 \begin{equation}
 U=U_{\rm s}+U_{e}+U_{\rm pion}+U_{\rm \nu}+U_{\gamma}. \label{eq2}
 \end{equation}

For the sake of simplicity, an SS was suggested to have a nearly uniform density from its center to the surface as the previous discussion mentioned. We may approximate an SS as a star with homogenous density of $\rho=3\rho_{0}$. If the average number of quarks in a strangeon is 10, then the average mass of strangeons is about 3 times higher than that of a nucleon, $m_{\rm n}$. We can obtain the strangeon number density $n_{s}=\rho /3m_{\rm n}$ and the total number of baryons in a star is $N_{\rm s}=V n_{\rm s}$. In our model, strangeons behave as classical particles (\citealt{Xu+2003}), then internal energy of thermal strangeon excitation strangeons in an SS is
\textbf{
\begin{equation}
U_{\rm s}=\frac{3}{2}nk\cdot4\pi\int_{0}^{R}r^{2}T_{r}dr, \label{eq3}
\end{equation}
}
where $T_{\rm r}$ is stellar temperature at point with distance $r$ from center of sphere.

Pions are mesons with zero spin. According to Bose-Einstein statistics, the average number of mesons in volume $V$ and with momentum $p$ and $p+dp$ is
\begin{equation}
  \frac{4\pi V}{h^{3}}p^{2}\frac{dp}{e^{\frac{\varepsilon -\mu}{kT}-1}}, \label{eq4}
\end{equation}
 where the relation between momentum $p$ and energy $\varepsilon$ is $\varepsilon^{2}=p^{2}c^{2}+m^{2}c^{4}$. Considering that, there are 3 kinds of pion. The internal energy of pions is
\textbf{
\begin{equation}\
U_{\rm pion}=3\cdot 4\pi\int_{0}^{R}\int_{140}^{\infty}r^{2}\frac{4\pi }{h^{3}}\frac{\varepsilon}{e^{\frac{\varepsilon -\mu}{kT_{r}}}-1}\frac{\varepsilon  \sqrt{\varepsilon^{2}-m^{2}c^{4}}}{c^{3}}d\varepsilon dr. \label{eq5}
\end{equation}
}

In the newborn strangeon star, the temperature is extra high that the collision frequency between particles is also very high. Then the system is almost at thermal equilibrium state. On the other hand, the time scale of reaction from neutrino to pion is much longer, thus the system is not at chemical equilibrium state. Therefore the chemical potential of pions and neutrinos is unequal. In this case, the chemical potential of pions can be approximately treated as pion's rest mass, $m\sim140$\,MeV, so the lower limit of integration in Eq.~\ref{eq5} is chosen to be $\varepsilon=140$\,MeV.

The exhaustive dynamic strangeonization process is unsettled, that a new-born strangeon star is isothermal  (i.e., temperature gradient negligible if turbulent convention dominates) or non-isothermal (i.e., temperature gradient significant) is uncertain, both of these assumptions should be considered. So we can get internal energy in both situations: isothermal proto-strangeon stars and non-isothermal proto-strangeon stars. In the case that a new-born strangeon star is a isothermal ball, $T_{\rm r}$ is $r$-independent and steady from center to surface. However, considering that heat transfer in early stage is mainly through neutrino diffusions, temperature gradient could exist in a proto-SS because neutrinos are opaque, as we prove in \S~\ref{sec2.2}. Then $T_{\rm r}$ should be a function of radius $r$, and we get the relation of $T_{\rm r}$ and surface temperature $T_{\rm s}$ as $T_{\rm r}\sim T_{\rm s}(\frac{R-r}{l})^{1/4} $, which is derived in \S~\ref{sec2.2}. With any given surface temperature $T_{\rm s}$, we can get corresponding internal energy by the temperature gradient relation.

In order to make a lower energy state of electrons, both NS model and SS model would go through a process to cancel the electrons by weak interaction. Considering that the number of electrons $N_{e^{-}}$ is generally around $10^{-5}$ of strangeon number $N_{\rm s}$, thus  $U_{e^{-}}$ could be ignored. We also ignore $U_{\nu}$ (and $U_{e^{+}}$) and $U_{\gamma}$ because these parts contribute little to the total internal energy, the specific calculations will be showed in \S~\ref{sec2.2}.

\begin{figure}
\centering
\subfigure[internal energy of isothermal strangeon stars]{
\label{figa} 
\includegraphics[width=2.65in]{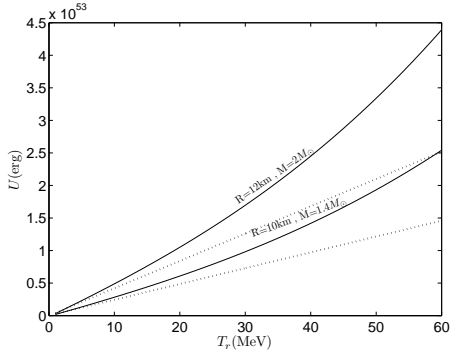}}
\hspace{0.2in}
\subfigure[internal energy of non-isothermal strangeon stars]{
\label{fig:subfig:b} 
\includegraphics[width=2.65in]{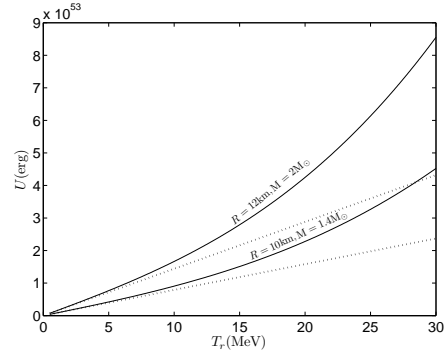}}
\caption{SSs' internal energy as a function of stellar temperature $T$. The black lines mean the total energy $U$, with mass $2M_\odot$ and $1.4M_\odot$ from top to bottom. The dash lines just stand for the corresponding $U_{\rm s}$. It is obvious that pions have a great influence on the total internal energy of SSs at high temperature.}
\label{Fig2} 
\end{figure}

Considering only the components which have dominated contributions to the internal energy, $U$ for both isothermal and non-isothermal cases are showed in Fig.~\ref{Fig2}. It is clear that $U$ of an SS with a different mass is around $10^{53}$\,erg, this result is consistent to the magnitude of binding energy. Making a comparison of the black line and dash line, which respectively corresponds to $U$ and $ U_{\rm s}$, we conclude that it is valid to consider pions as an important freedom degree of a newborn SS.
   \begin{figure}
   \centering
   \subfigure[]{
  \includegraphics[width=2.65in]{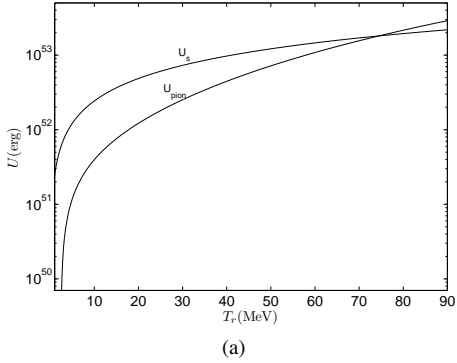}}
  \hspace{0.2in}
  \subfigure[]{
  \includegraphics[width=2.65in]{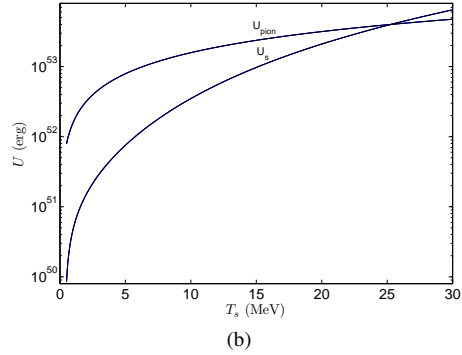}}
  \caption{A comparison between $U_{\rm s}$ and $U_{\rm pion}$ of an SS with $M=1.4M_{\odot}$, $R=10$ km. At the beginning, $U_{\rm pion}$ is almost the same order of magnitude with $U_{\rm s}$, even larger than $U_{\rm s}$ if the initial temperature is high. As $T$ drops down, pions cannot be excited and begin to decay quickly. As we can see from this plot, $U_{\rm pion}$ reduces rapidly and can be ignored when $T$ drops below several MeV.}
   \label{Fig3}
   \end{figure}
Our results, shown in Fig.~\ref{Fig3}suggest that pions share almost half of the gravitational binding energy at an initial temperatures which are roughly $40\sim50$ \,MeV for isothermal case and 10 MeV for non-isothermal case, according to Eq.~\ref{eq5}. When the new-born SS cools down, pions will decay rapidly because they are unstable. Then large amount of neutrinos will be released by pion decay. Therefore pions would be insignificant, and it would be unnecessary to be considered during the later thermal evolution when $T$ decreases to several MeV, as Fig.~\ref{Fig3} shows.

\subsection{Neutrino emissivity of proto-strangeon star}
\label{sec2.2}

Whether it is in neutron stars or strangeon stars, neutrino emission is similar to photon radiation in early stage (\citealt{Bethe+Wilson+1985, Janka+etal+1989+a, Janka+etal+1989+b}), just like blackbody radiation. It is well known that neutrinos are less-massive particles, they pass through common substances almost freely because they are only affected by weak interaction with extremely small scattering cross-sections. But neutrinos produced in the newborn SS can hardly escape freely from inside to surface because strangeon matter is so dense that the neutrinos are trapped and matter in SSs becomes opaque.

Generally, absorption and scattering is the main mechanism of neutrino opacity. For the case of free quark matter, absorption processes ($d+\nu_{e}\rightarrow u+e^{-}$, $s+\nu_{e}\rightarrow u+e^{-}$) could play a more significant role in determining the mean free path of the neutrinos than scattering processes ($q+\nu \rightarrow q+\nu, q=n,p$). However, in the normal nucleus matter, mean free path of absorption and scattering processes are almost at the same order (\citealt{Iwamoto+1982}). In addition, the $\beta$ equilibrium should has been reached when a new-born SS is form, thus a significant absorption possible could not be kinematically allowed in the strangeon star. So we consider only the scattering process.

\begin{figure}
  \centering
  \includegraphics[width=10cm]{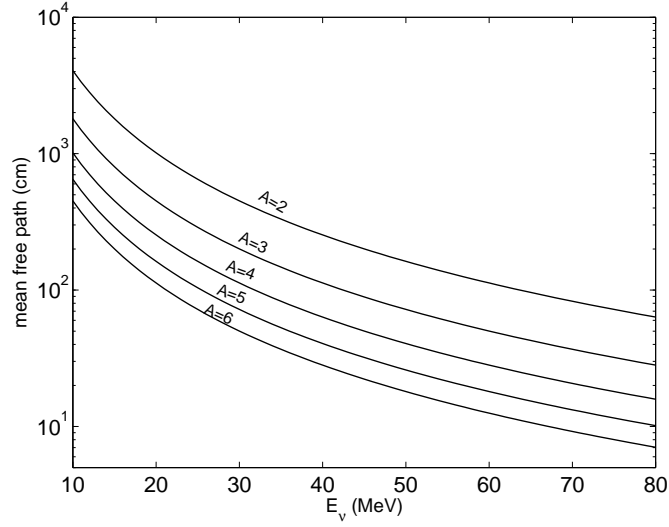}\\
  \caption{The mean free path of neutrinos in strangeon star. We are taking the range of $E_{\nu}$ from 10 MeV to tens of MeV as typical supernova neutrino energies. When a strangeon consists more quarks with larger $A$ (such as $A=6$ means a strangeon with 18 quarks), it corresponds to a shorter mean free path. Then it is hard to escape from SSs for neutrinos with high energy.}
  \label{Fig4}
\end{figure}

Considering that a strangeon is a cluster with a certain number of quarks, it is convenient to take strangeons as special nucleons with strangeness when scattering with neutrinos. In Weinberg's weak interaction theory, Freedman obtained the differential cross section for neutrino-nucleus scattering (\citealt{Freedman+1974}), which reads
\begin{equation}
\frac{d\sigma}{dq^{2}}=\frac{G^{2}}{2\pi}a_{0}^{2}A^{2}e^{-2bq^{2}}(1-q^{2}\frac{2ME_{\nu}+M^{2}}{4M^{2}E^{2}_{\nu}}), \label{eq6}
\end{equation}
where $G$ is the conventional Fermi constant: $G=1.015\times10^{-5}m_{p}^{-2}$, $\theta_{W}$ is the Weinberg angle, $a_{0}=-\sin^{2}\theta_{W}$ ($\sin^{2}\theta_{W}=0.23\pm0.015$), $A$ is the nucleon number of the target nucleus and $b$ is related to the target particle radius $r$ by $b=\frac{1}{6}r^{2}\approx 4.8\times 10^{-6}A^{2/3}$. Parameter $q^{2}$ is the squared momentum transfer. Considering the fact that neutrinos almost have no interaction with the targets, we just take $q\ll E_{\nu}$, then for supernova neutrinos, the part in brackets is approximately unity with energy $E_{\nu}\sim10$\,MeV and strangeons $M\sim3\times10^{3}$\,MeV. Integrating Eq.~\ref{eq6}, one has
\begin{equation}
 \sigma \approx0.03\times\sigma _{0}A^{2}(\frac{E_{\nu}}{m_{e}c^{2}})^{2}, \label{eq7}
\end{equation}
where $\sigma _{0}=1.7\times10^{-44}$ $\rm cm^{2}$. In this case, we can consider a strangeon to be a cluster with baryon number $A$,  and the mean free path of neutrinos in a proto-SS will be $l=(n_{\rm sc}\sigma (\nu A))^{-1}$, as we can see in Fig.~\ref{Fig4}.

For a new-bown hot SS, the heat transfer before solidification is mainly through neutrino diffusions. As we calculated above, neutrinos in the prpto-SS are opaque, then thermal energy energy delivery will be blocked. Therefore, temperature gradient could exist in a proto-SS. However, the dynamic process of strangeon star formation is still uncertain, effect of temperature gradient may not be ignored. In this section, we take the temperature gradient into consideration and recalculate the cooling process as a contract to the isothermal case.

Gudmundsson et.al. (1982) had researched the temperature differences between the core and surface. On the surface of a neutron star, photon luminosity can be expressed by $L=4\pi\sigma T_{s}^{4}=f(\kappa,T_{c})$, where $T_{s}$  and $T_{c}$ are surface and central temperature, $f(\kappa,T_{c})$ is a function related to structure and statement equations of neutron stars. Similarly, we can get a temperature gradient relation in a roughly way.

For a proto-SS, we set the internal temperature and surface temperature as $T_{r}$ and $T_{s}$, where $r$ is the distance to the central of the proto-SS, and radius of the star can be set as $R$. In the ideal situation, if matter is transparent to neutrinos, idea elapsed time for neutrinos at $r$ position is $t_{1}\sim (R-r)/c$, and in position $R$ luminosity is the order $R^{2}T_{r}^{4}$. In this case, the whole star share same temperature $T_{r}$. But, in fact, strangeon matter is opaque to neutrinos at early stage. Neutrinos can escape only after numbers of collisions, which can be thought as ``random walking'' process. Then number of collisions is $N\sim (R-r)^{2}/l^{2}$ where $l$ is the mean free path of neutrinos in a proto-SS.  So the virtual elapsed time is $t_{2}\sim Nl/c\sim (R-r)^{2}/(lc)$, and the virtual luminosity is the order $R^{2}T_{s}^{4}$, so that the process of radiative diffusion has been slow down the rate at which energy escaped the proto-SS by a factor $t_{2}/t_{1}\sim(R-r)/l$.

By the energy conservation, the elapsed-time delay leads to temperature gradients in  proto-SSs. Internal luminosity, of order $R^{2}T_{r}^{4}$ , is reduced to the surface luminosity, of order $R^{2}T_{s}^{4}$. Thus $(T_{r}/T_{s})^{4}\sim(R-r)/l$, and we get the an estimate of the temperature relation as
 \begin{equation}
 T_{r}\sim T_{s}(\frac{R-r}{l})^{1/4}  .\label{eq21}
\end{equation}

The mean free path $l$ is just ($10^{-4}\sim 10^{-3})$  $R$, which means that only in a very thin sphere shell on the surface can neutrinos be emitted out freely, so we can describe this escaping process as bulk emission in comparison with photon radiation which is also considered as surface emission. Thickness of the emission shell can be regarded as the mean free path $l$, and we take $l=10^{3}$\,cm. That is to say, neutrinos below the shell cannot escape immediately. They are trapped in the star and form the so-called `` neutrinosphere''. The opaque neutrino emission field presented as surface emission is on the interface below the free emission shell, just like photo blackbody radiation. In other words, the total luminosity of neutrinos is composed of two parts, bulk neutrino emission luminosity $L_{\rm b\nu}$ and surface neutrino emission luminosity $L_{\rm s\nu}$. Then we calculate both of them to get the entire neutrino emission luminosity.

In high temperature (such as the case of a new-born strangeon star or neutron star), via pair annihilation ($\gamma +\gamma \leftrightarrow e^{\pm}\rightarrow {\nu}+\overline{\nu}$), which is in the frame work of Weinberg-Salam theory, is the dominating form of neutrino energy-loss rates than photo-, plasma and bremsstrahlung process(\citealt{Itoh+etal+1989}). As an indication, we consider only this mechanism since we do not exactly know the neutrino energy-loss rate of the strangeon matter. We often use emissivity in unit volume to calculate neutrino emission energy (\citealt{Braaten+etal+1993}). The emissivity of neutrino with high temperature ($T>1$\,MeV) from Itoh et al.(1989) is
\begin{equation}
\varepsilon _{\rm pair}=1.809(1+0.104q_{pair})f(\lambda)g(\lambda)e^{-2/\lambda} \rm erg\cdot \rm s^{-1}\cdot \rm cm^{-3}, \label{eq8}
\end{equation}
and
$$q_{pair}=(10.7480\lambda ^{2}+0.3967\lambda ^{0.5}+1.0050)^{-1}[1+(\rho/\mu_{e})(7.692\times 10^{7}\lambda^{3}+9.715\times10^{6}\lambda^{0.5})^{-1.0}]^{-0.3},$$
$$g(\lambda)=1-13.04\lambda ^{2}+133.5\lambda ^{4}+1534\lambda ^{6}+918.6\lambda^{8},$$
$$f(\lambda)=\frac{(6.002\times10^{19}+2.084\times10^{20}\xi+1.872\times10^21\xi^{2})e^{-4.9924\xi}}{\xi^{3}+1.2383/\lambda-0.4141/\lambda^{2}}$$
where $\lambda =\frac{T}{5.9302\times 10^{9}K}, \xi=[\rho\mu_{e}^{-1}/(10^{9}\rm g\cdot \rm cm^{-3})]^{1/3}\lambda^{-1}$. In SS model, the number of electrons per baryon is $<10^{-4}$ than quarks, so we choose the electron mean molecular weight $\mu_{e}=10^{5}$ in the following calculations. Therefore the bulk neutrino emission luminosity is
\begin{equation}
L_{b\nu}=4\pi R^{2}l\varepsilon _{\rm pair}. \label{eq9}
\end{equation}

Next we consider the surface emission which is similar to the blackbody radiation. The `` neutrinosphere'' below the thin free emission shell can be thought as neutrino radiation field with the radius  $R-l\approx R$. In Fermi-Dirac statistics, emission intensity of neutrinos is
\begin{equation}
I_{\nu}=\frac{\varepsilon _{\nu}}{c^{2}h^{3}}\frac{1}{e^{(\varepsilon_{\nu}-\mu_{\nu})/kT}+1}, \label{eq10}
\end{equation}
where $\varepsilon _{\nu}$ is neutrino energy, and the chemical potential $\mu_{\nu}=0$. In the radiation field, energy density is
\begin{gather}
  u_{\nu}=\frac{4\pi}{c}\int_{0}^{\infty}I_{\nu}d\varepsilon_{\nu} \notag\\
  =\frac{4\pi (kT)^{4}}{(hc)^{3}}F_{3}, \label{eq11}
\end{gather}
where $F_{3}$ is the Fermi integral. The internal energy of a new born SS is reviewed in \S~\ref{sec2.1}, we can then use Eq.~\ref{eq11} to estimate $U_{\nu}\sim 4/3\pi R^{3} u_{\nu} $ $\sim 10^{48}$ erg, and $U_{\gamma}$ should be smaller, thus we ignore these two components of the total internal energy in Eq.\ref{eq2}. Like photons, the flux of the neutrino radiation is $\frac{c}{4}u_{\nu}$. Considering 3 flavors of neutrinos and their anti-particles, then it yields
\begin{equation}
L_{\rm s\nu}=6\cdot4\pi R^{2}\sigma_{v}T^{4}, \label{eq12}
\end{equation}
where $\sigma_{v}\approx 14.88\times 10^{-5}$ erg$\cdot$ cm$^{-2}\cdot$ s$^{-1}\cdot $K$^{-4}$ based on Eq.~\ref{eq11}.

Another form of radiation for the cooling process is photon radiation. We regard this part as blackbody radiation,
 \begin{equation}
 L_{\gamma}=4\pi R^{2}\sigma T^{4}. \label{eq13}
 \end{equation}

The main dissent in Eq.~\ref{eq13} is that $T$ is the so-called effective temperature $T_{e}$ in other models. In these models, neutron stars have complex structures, and are usually with a crust on the surface (\citealt{Pethick+Ravenhall+1995}) which generates a temperature gradient from the center to the surface, and $T_{e}$ is generally much lower than 10 MeV, which is the order of surface temperature of a bare new-born SS. In this case, $T$ in Eq.~\ref{eq13} is the same as in Eq.~\ref{eq12}, then the energy released by photons is about $10^{52}$\,erg in our calculation. This energy is more than the total energy needed to drive a supernova (usually $1\%$ of the gravitational binding energy). \cite{Chen+etal+2007} did some specific research on the huge energy carried out by photons and found that a supernova may actually be driven by photons.

 \subsection{Thermal evolution of proto-strangeon star with solidification}
 \label{sec2.3}

The above calculations are aimed at exploring the rapid cooling stage through releasing neutrinos. Firstly we should confirm the internal energy (which equals the binding energy in \S~\ref{sec2.1}) and initial temperature of a new born SS. A simple approximate ``empirical formula'' describes  $E_{\rm bind}$ well at $M>0.5M_\odot$ (\citealt{Lattimer+Yahil+1989}), and we use it to estimate the binding energy of an SS as
\begin{equation}
E_{\rm bind} \simeq 1.5\times10^{53}(M/M_\odot)  \rm erg.  \label{eq14}
\end{equation}
From Eq.~\ref{eq14} and the parameters which are $M=1.4M_\odot$, $M=2M_\odot$ we can estimate that if an SS is born in SN 1987A, the total thermal energy at the beginning is around $2.1\times 10^{53}$\,erg, $3\times10^{53}$\,erg, Equaling the binding energy to the internal energy $U$ in Eq.\ref{eq2}, the initial temperatures are respectively $T_{\rm r}=52.9$\,MeV, 50.7\,MeV(isothermal new-born SSs) and $T_{\rm s}= 17.8$\,MeV, 16.4\,MeV (non-isothermal new-born SSs). So we take $T_{r}=50$\,MeV, $T_{\rm s}=18$\,MeV for all numerical calculations in this paper.

The internal energy loss rate of an SS at the beginning is
 \begin{equation}
 -\frac{dU}{dt}=L_{\rm b\nu}+L_{\rm s\nu}+L_{\gamma}. \label{eq15}
 \end{equation}

\begin{figure}
   \centering
   \includegraphics[width=10.0cm, angle=0]{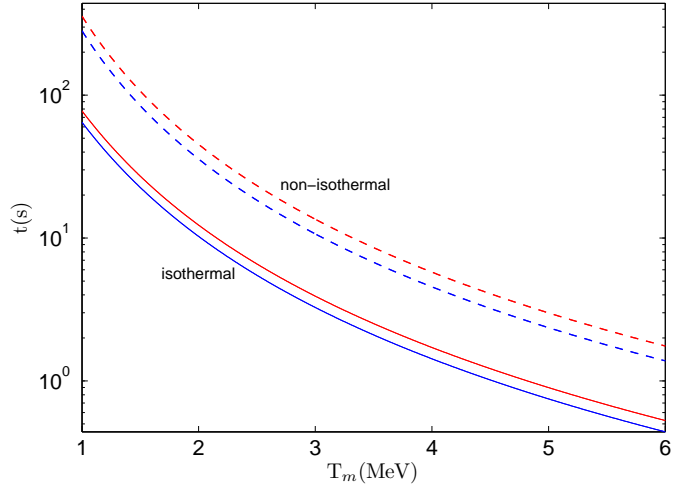}
   \caption{Time scale of the constant-temperature stage of phase transition process. Among these curves, the red ones have parameter as$M=2M_{\odot}$, $R=12$ km, and the blue ones correspond to $M=1.4M_{\odot}$, $R=10$ km. The dashed curves are the time scale of proto-SSs with temperature gradient, and the solid cures below are isothermal proto-SSs. It can be seen that the lasting time of homothermal stage is highly sensitive to melting temperature. }
   \label{Fig5}
   \end{figure}

The evolution which Eq.\ref{eq15} represents lasts during the entire cooling process of normal NSs or SQSs. This process, which is represented by a $T-t$ relation curve with temperature $T>T_{\rm m}$, is shown in Fig.~\ref{Fig6} with different $T_{\rm m}$.  As mentioned in \S~\ref{sec1} , an SS would go through a phase transition from liquid to solid, and this cooling process will not last long. We have declared that the strangeon behaves like classical particles, therefore strangeon matter would be localized in crystal lattice if the stellar temperature reaches its melting temperature $T_{\rm m}$ which has a range of 1\,MeV $<T_{\rm m} <$6\,MeV (\citealt{Xu+2003, Dai+etal+2011, Lai+etal+2013}). The cooling star will remain a stable temperature ($T=T_{\rm m}$) for a while, and during the stage from liquid to solid, the latent heat would be released through thermal emission. We obtain the time scale of the constant-temperature stage from
\begin{equation}
E'=(L_{\rm b\nu}+L_{\rm s\nu}+L_{\gamma})t. \label{eq16}
\end{equation}
where $E'$ is the latent heat. To estimate the latent heat, we need to know the state of cold quark matter and interactions between strangeons. Lai and Xu (\citealt{Lai+etal+2009}) used the Lennard-Jones potential to describe the
interaction between strangeons and gave the depth of the potential $V\sim 100$MeV. Then the latent heat released by each strangeon can be written as $\varepsilon_{s}=fV$, where $f$ is the ratio of potential to melting heat. Based on this work, considering that strangeons are non-relativistic and the interaction is similar to common substances, it is reasonable to estimate $f$ to be $0.01\sim0.1$, which is the radio for most common substances. Then the energy released by each strangeon in the liquid to solid phase is $\varepsilon_{s}\sim1-10$ MeV for estimation (\citealt{Dai+etal+2011}). The total latent heat of SSs can be written as
\begin{equation}
E'=N_{s}\varepsilon_{s}, \label{eq17}
\end{equation}
and the results are $3\times10^{51}$erg,$5\times10^{51}$erg with the corresponding mass $1.4M_{\odot}$, $2M_{\odot}$ if $\varepsilon_{s}\sim1$ MeV for estimation.
When temperature cools down to the melting temperature which we choose 3 MeV here, the current internal energy $U'$ respectively are $7.2\times 10^{51}$erg, $1.3\times10^{52}$erg. From the comparison of latent heat and current internal energy, it is obvious that most part of the current internal energy is released in the constant-temperature stage. Due to uncertainty of many parameters of latent heat, such as potential, the ratio of potential to melting heat, we use $U'$ to replace $E'$ in Eq.~\ref{eq16} to get the time scale of the latent heat releasing process, and results are shown in Fig.~\ref{Fig5}.

Considering the whole thermal evolution, the lasting time of latent heat releasing process is represented by part of the $T-t$ curve, as shown in  Fig.~\ref{Fig6} and Fig.~\ref{Fig7} with different melting temperature.

After this homothermal stage, proto-SSs crystallize immediately and finally become solid state. Residual internal energy for SSs in a solid state can be wrote as
\begin{equation}
U_{\rm re}=\int C_{V}dT, \label{eq18}
\end{equation}
where heat capacity $C_{V}$ comprises of lattice structure component $C_{V}^{l}$ and electron component $C_{V}^{e}$, and then $C_{V}=C_{V}^{l}+C_{V}^{e}$. Because of the small mount of electrons, $C_{V}^{e}$ can be ignored (\citealt{Yu+Xu+2011}). Pions will not be taken into consideration in this part. As mentioned in \S~\ref{sec2.1}, Fig.~\ref{Fig3} showed that when the $T$ drops down to several MeV, pions decay quickly.

Debye model is thought to be quite an appropriate method to estimate the specific heat of solid state SSs (\citealt{Yu+Xu+2011}). If a solid medium consists of strangeons, the specific heat is
\begin{equation}
C_{V}^{l}=N\cdot\frac{12\pi^{4}}{5}k(\frac{T}{\theta_{D}})^{3}, \label{eq19}
\end{equation}
where $\theta_{D}=\hbar (\overline{C_{s}}k_{D})/k$ is Debye Temperature in which the average sound speed of SSs is $\overline{C_{\rm s}}\sim c$, and D$k_{D}=(6\pi^{2}n_{\rm s})^{1/3}$ is Debye wave number where $n_{\rm s}$ is number density of strangeons. Because the number density of particles of an SS is extremely high when compare with common substances, Debye Temperature for SSs is as high as $10^{12}$ K.
After crystallization, thermal evolution is represented as
\begin{equation}
 -C_{V}^{l}\frac{dT}{dt}=L_{\rm b\nu}+L_{\rm s\nu}+L_{\gamma}. \label{eq20}
\end{equation}

Because of the relatively small heat capacity of solid SSs in Eq.\ref{eq20}, temperature drops down sharply, and it is shown in Fig.~\ref{Fig6}. A sharp decrease of temperature will lead to an extremely small flux of neutrinos, which means the violent releasing of neutrinos, i.e. neutrino burst, will cut off after phase transition.

Combination of Eq.\ref{eq15}, \ref{eq16}, \ref{eq20} is the representation of the whole cooling down process of proto-SSs, and the whole process is shown in Fig.~\ref{Fig6},~\ref{Fig7}. Temperature decreases in this process is mostly resulted by neutrino emission, and this process corresponds to the detected neutrino burst. The specifics will be discussed below.

\section{ the neutrino burst of SN 1987A in a strangeon star model}
\label{sec3}

Neutrino burst is one of the astronomical phenomena during the fast cooling stage of a newborn compact star in a supernova. If an SS is born in SN 1987A, the neutrino burst should be explained by SS model. In this section, we test the thermal evolution of proto-SSs we studied in \S~\ref{sec2.2} and \S~\ref{sec2.3}, by SN 1987A neutrino burst.

\subsection{Neutrino burst events in SN 1987A}
\label{sec3.1}
The SN 1987A neutrino burst was detected by 3 detectors (\citealt{Hirata+etal+1987, Hirata+etal+1988, Bionta+etal+1987, Bratton+etal+1988, Alekseev+etal+1987, Lore+etal+2002}), and all neutrino events observed are listed in Tab.~\ref{tab1}.

With different energy thresholds, these 3 detectors detected different numbers of neutrino events. Energy threshold of Kamiokande-\uppercase\expandafter{\romannumeral2} is 7.5\, MeV, and in early data, events K6, K13, K14, K15, K16 were not included. However, these 5 neutrinos were picked up from neutrino background and were included in this neutrino burst in subsequent analyses (\citealt{Lore+etal+2002, Vissani+2015}). The energy thresholds of the other detectors are $15\rm MeV$ for IMB and $10\rm MeV$ for Baksan. With lower energy threshold, Kamiokande-\uppercase\expandafter{\romannumeral2} could detect much more events than the other two, as shown in Tab.~\ref{tab1}.

In addition, we can hardly determine when the neutrino burst begins exactly. Considering the uncertainty of universal time, the first event observed by Kamiokande-\uppercase\expandafter{\romannumeral2}, IMB, Baksan occurred at these corresponding specific times 7:34:35 UT$\sim$7:36:35 UT, 7:35:40.95 UT$\sim$7:35:41.05 UT, and 7:35:18 UT$\sim$7:36:14 UT (\citealt{Aglietta+etal+1990}). In this situation, a separate analysis of these 3 groups of data may be more accurate when researching a time-dependent physical process, such as $T-t$ evolution of proto-NSs or proto-SSs.

\begin{table}
\bc
\begin{minipage}[]{100mm}
\caption[]{Properties of the detected neutrino burst events in SN 1987A. Events K1,K2...K16 are detected by Kamiokande-\uppercase\expandafter{\romannumeral2}, I1,I2...I8 and B1,B2...B5 were recorded by IMB and Baksan respectively. Relative time here means the starting moment of each detector's first event, do not represent the absolute starting time of neutrino burst. ~\label{tab1}}\end{minipage}
\setlength{\tabcolsep}{1pt}
\small
 \begin{tabular}{c c c | c c c}
  \hline\noalign{\smallskip}
    Detector   &   Relative time  &  Energy  &  Detector  &  Relative time  &  Energy\\
    &(s) & (MeV)  &  &(s) & (MeV)\\
    \hline\noalign{\smallskip}
    K1&0&20.0$\pm$2.9&I1&0&38$\pm$7\\
    K2&0.107&13.5$\pm$3.2&I2&0.412&37$\pm$7\\
    K3&0.303&7.5$\pm$2.0&I3&0.650&28$\pm$6\\
    K4&0.324&9.2$\pm$2.7&I4&1.141&39$\pm$7\\
    K5&0.507&12.8$\pm$2.9&I5&1.562&36$\pm$9\\
    K6&0.686&6.3$\pm$1.7&I6&2.684&36$\pm$6\\
    K7&1.541&35.4$\pm$8.0&I7&5.010&19$\pm$5\\
    K8&1.728&21.0$\pm$4.2&I8&5.582&22$\pm$5\\
    K9&1.915&19.8$\pm$3.2&&&\\
    K10&9.219&8.6$\pm$2.7&&&\\
    K11&10.433&13.0$\pm$2.6&&&\\
    K12&12.439&8.9$\pm$2.9&B1&0&12.0$\pm$2.4\\
    K13&17.641&6.5$\pm$1.6&B2&0.435&17.9$\pm$3.6\\
    K14&20.257&5.4$\pm$1.4&B3&1.710&23.5$\pm$4.7\\
    K15&21.355&4.6$\pm$1.3&B4&7.687&17.5$\pm$3.5\\
    K16&23.814&6.5$\pm$1.6&B5&9.099&20.3$\pm$4.1\\
    \noalign{\smallskip}\hline
\end{tabular}
     \ec
\end{table}

\subsection{To understand the neutrino burst events in strangeon star cooling model}
\label{sec3.2}

When discussing the time-dependent cooling process of SSs and testing the model with the observed events, it is obviously unsuitable to make a combined analysis about all data together. Because of the aforementioned uncertainty of the universal time between 3 laboratories, the exact moments of the first event of the 3 detectors were uncertain. For this reason, data from 3 different timelines cannot be analysed by sharing a common starting-time. For the sake of preciseness and objectiveness, we finally chose Kamiokande-\uppercase\expandafter{\romannumeral2}'s events as the optimal sample, without a combined analysis to IMB and Baksan.

The relation between neutrino energy and stellar temperature can be derived from neutrino distribution function $f=E_{\nu}^{2}/(1+ exp(E_{\nu}/T))$, where $T$ is the temperature of SSs in our model, then the mean energy can be obtained as (\citealt{Janka+etal+1989+b})
  \begin{equation}
  \langle E_{\nu}\rangle=\frac{\int_{0}^{\infty}E_{\nu}fdE_{\nu}}{ \int_{0}^{\infty}fdE_{\nu}}\approx3.15T. \label{eq21}
  \end{equation}
We use relations $E_{\nu}\sim3.15T$ to represent the neutrinos' energies with SS's temperature, and understand these 15 time-dependent events (without event K1) together with the $T-t$ evolution.
In this case, the theoretically cooling curves of proto-SSs calculated in \S~\ref{sec2} could be tested by the observed 15 neutrino events.

\begin{figure}
  \centering
  \subfigure[Cooling process for isothermal proto-strangeon stars]{
  \includegraphics[width=2.7in]{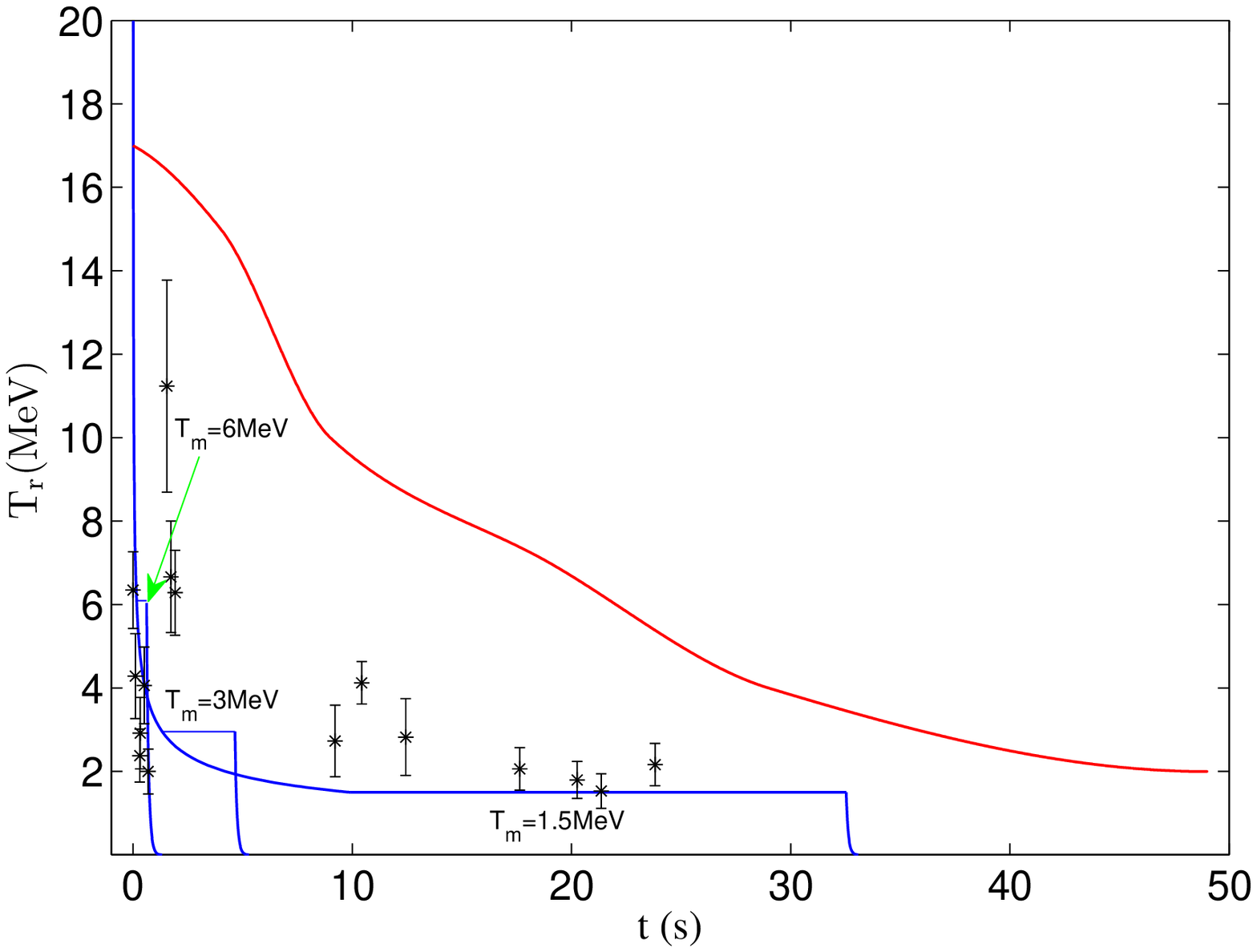}}
  \hspace{0.05in}
  \subfigure[Cooling process for non-isothermal proto-strangeon stars]{
  \includegraphics[width=2.7in]{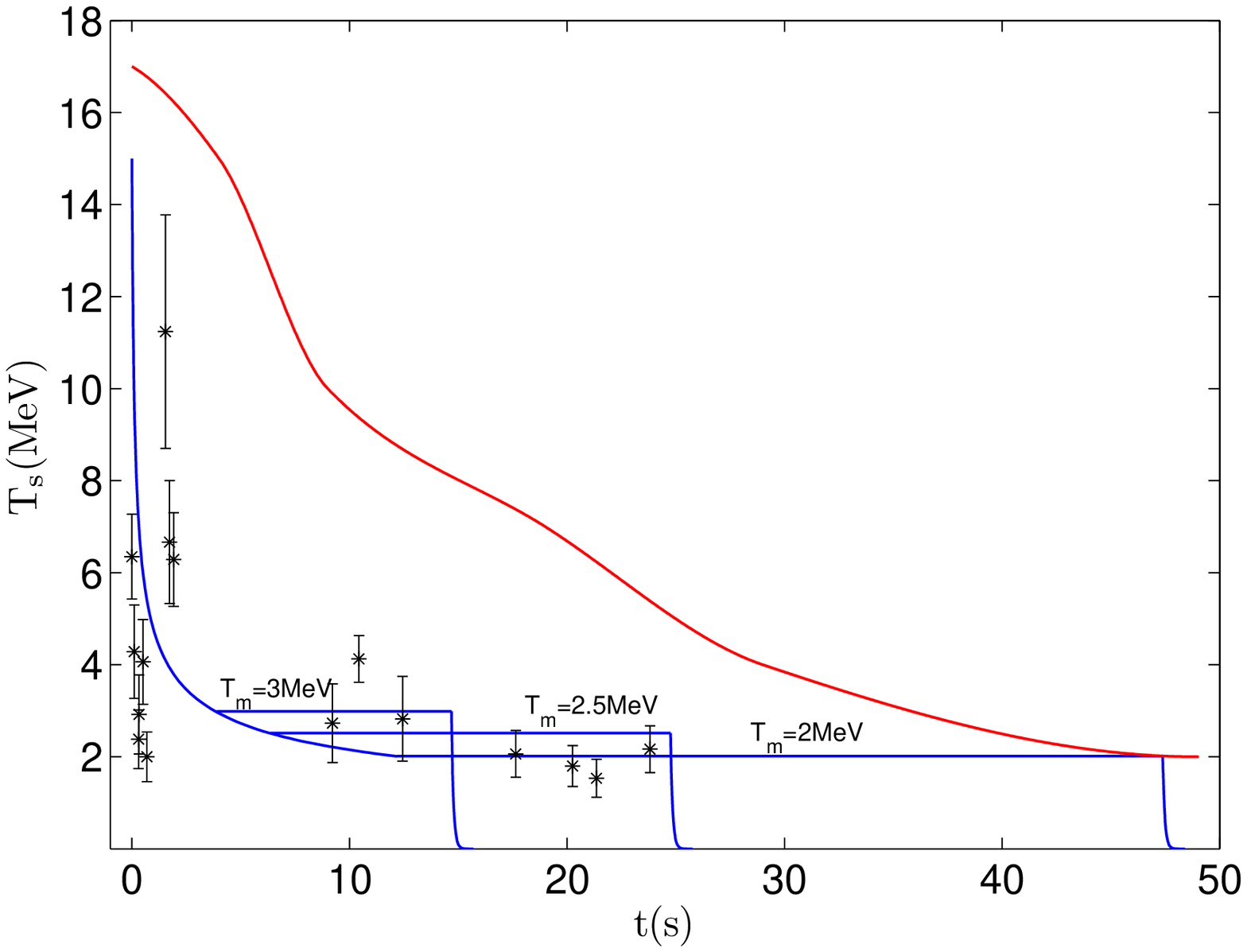}}
  \caption{This is the $T$-$t$ relation of a proto-SS with $M=1.4M_{\odot}, R=10$ km. Both in figure (a) and (b), the upper red curve is the cooling process of a normal proto-NS (with mass $M'=1.4M_{\odot}$) taken from Pons (\citealt{Pons+etal+1999}). The 15 black dots with error bars are 15 neutrino events, of which the neutrino energy has been represented by proto-SSs' temperature with a relation of $E_{\nu}=3.15 T$. It can be seen that $T$ in a proto-SS drops down rapidly in early stage, and then the star will keep homothermal during phase transition, as the straight lines indicate. After phase transition, solid SSs cool down more drastically than ever before. In this case, consequently, the emission intensity decreases quickly, and hence leading to a cut-off of this neutrino burst. On the contrast, normal NSs cool down smoothly all the time, and having no interceptive indication during the SN 1987A neutrino burst. It indicates that if the proto-SS can be thought as isothermal ball, melting temperature $T_{\rm m}\sim 1$ MeV coincides well with the neutrino burst events. Taking temperature gradient into consideration, $T_{\rm m}\sim 2$ could be fine. Both of these two situations can roughly fit the neutrino bust. }
  \label{Fig6}
\end{figure}

\begin{figure}
  \centering
  \subfigure[]{
  \includegraphics[width=2.70in,angle=0]{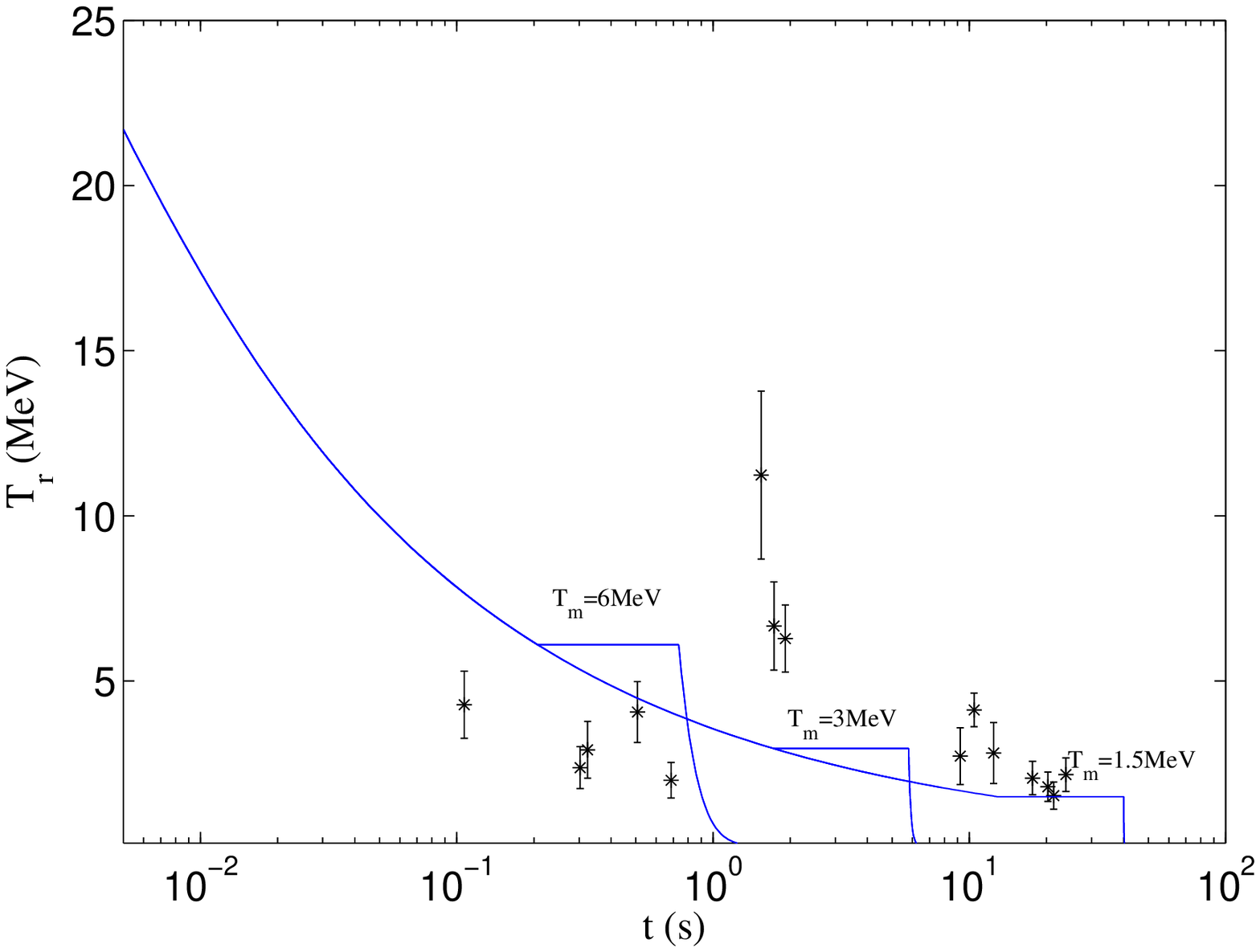}}
  \hspace{0.1in}
  \subfigure[]{
  \includegraphics[width=2.70in,angle=0]{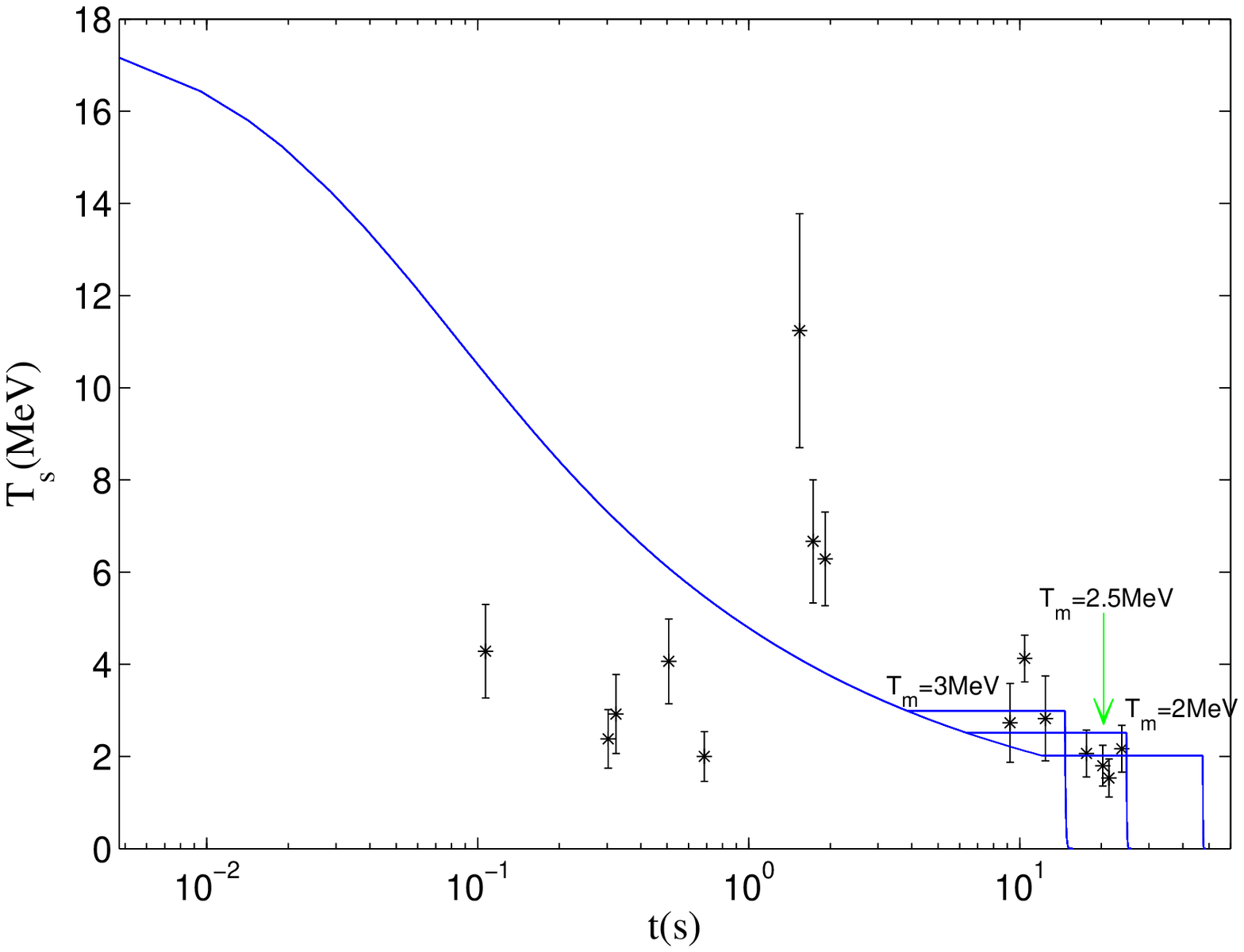}}
  \caption{$T-t$ relations of SSs for $M=2M_{\odot}, R=12\rm km$. The curves in (a) and (b) are presented by taking logarithm of $t$, as a more detailed presentation of the early cooling stage.}
  \label{Fig7}
\end{figure}

To the $T-t$ curves, we take melting temperature to be $T_{\rm m}$=6\,MeV,3\,MeV, 1.5\,MeV for isothermal SSs, and  $T_{\rm m}$=3\,MeV,2.5\,MeV, 2\,MeV for non-isothermal SSs. Each $T_{\rm m}$ corresponds to a different lasting-time of the phase transition. For comparison, we give the cooling curve of a normal proto-NS with the mass $1.4 M_{\odot}$ at the same stage, the results are shown in Fig.~\ref{Fig6}. And in Fig.~\ref{Fig7} we also present another two $T-t$ relations which corresponds to SSs with different masses. Because the normal NS model cannot support compact stars (not including a black hole) more than 2 $M_{\odot}$, there is no comparison in Fig.~\ref{Fig7}.

It appears that there is almost no difference between $T-t$ relations for $M=1.4M_{\odot}, R=8\rm km$. It indicates that a neutrino burst is a good way to examine the different pulsar-like object models, but not a good way to get information about the $M-R$ relation. However, all these results show that our solid SS model with a melting temperature around 1 MeV can reproduce the neutrino burst which can be well explained by observation.

  \section{Conclusions and Discussions}
  \label{sec4}

In this paper, comprehensive calculations are made on the entire thermal evolution of a newborn SS, including its thermal energy, radiation, and phasing transition. Our conclusions are as following. (1) Pion excitation could greatly contribute to the internal energy, as shown in Fig.~\ref{Fig3}. The total thermal energy of pions and strangeons are in accordance with the fundamental core collapse theories of a massive star. (2) The theoretically time-dependent temperature evolution of SS model, both isothermal one and the non-isothermal one,  coincide well with the SN 1987A neutrino burst when compared with the normal neutrino-driven NS model, as  indicated in  Fig.~\ref{Fig6}, Fig.~\ref{Fig7}. It is worth noting that an obvious cut-off of the neutrino burst after liquid-solid phase transition occurs in our model. The cut-off time becomes longer if the melting temperature is lower, and/or if the temperature gradient is more significant. A long neutrino burst during explosion would not be good for a successful supernova in normal neutron star model, but does not matter in the strangeon star model because of explosion photon-driven rather than neutrino-driven (\citealt{Chen+etal+2007}).

The characteristic cut-off of the neutrino burst in the model indicates that almost none supernova neutrinos can be detected after the liquid-solid phase transition, i.e., the solidification of strangeon matter. An dramatic low-flux neutrino emission (with terminated neutrino energy around 3 MeV) will make the detector hardly get consecutive neutrino events. However, in case of NS models without neutrino cut-off, a detector could be able to detect a large number of supernova neutrinos continuously even the neutrino energy is lower than 3 MeV. Therefore, it would be a way to test those two kinds of models by future advanced neutrino detectors.
New neutrino experiments are already underway with many planned for the near future. For example, JUNO (Jiangmen Underground Neutrino Observatory) could record more than thousands of neutrino events during a supernova like SN 1987A (\citealt{An+etal+2016}). The more events can be detected, the more information is available for the compact remnant, and we are looking forward to detecting supernova neutrino burst as well as to testing the models.

However, in the extremely early stage of the cooling process, which corresponds to the high temperature part and the time-dependent temperature evolution in early stage (before phase transition), some aspects need further discussions. Certainly less neutrino event was recorded in this very early stage than later cooling process, as shown in Tab.~\ref{tab1} and Fig.~\ref{Fig6}. We will discuss this issue on both technical and theoretical points of view as following.

Technically, each detector of the three has a limit to its trigger rate, which makes it hard to record the continual trigger at a millisecond interval. For example, the IMB detector is dead for about 35 ms after each trigger, which is of the same order in the case of Kamiokande-\uppercase\expandafter{\romannumeral2} (\citealt{Bionta+etal+1987}). What is more, the total number of photoelectrons per event in the Kamiokande-\uppercase\expandafter{\romannumeral2} photomultiplier tubes (PMT's) had to be less than 170, corresponding to a maximum electron energy with 50 MeV (\citealt{Hirata+etal+1988}).
In addition, due to the uncertainty of universal time between those detectors (K, I and B in Table 1), we cannot adopt all of the data at the same initial time, and therefore Kamiokande events are shown in Fig.~\ref{Fig6}.

Theoretically, we have not considered the interaction between neutrinos and the circumstellar matter of a newborn SS.
The neutrino mean free path could be much smaller than the length scale if one takes the enclosed mass of core collapse supernova as high as $0.4 M_{\odot}$ (in millisecond after supernova explosion, \citealt{Roberts+etal+2016}).
One may then expect that the time of arrival and spectrum of supernova neutrino should be modified if this circumstellar matter effect is included.

This work is supported by the National Natural Science Foundation of China (11673002, U1531243 and 11373011) and the Strategic Priority Research Program of CAS (No.XDB23010200).

\bibliographystyle{raa}
\bibliography{ms_ref}

\end{document}